\documentclass{rspublic}

\usepackage{graphicx}

\begin{document}

\title[Anomalous Exponent of Photoconduction]{Hot-Carrier Model for an Anomalous Exponent of Photoconduction}

\author[M. Wilkinson]{Michael Wilkinson}

\affiliation{Department of Mathematics and Statistics,
The Open University, Walton Hall, Milton Keynes, MK7 6AA, England}

\label{firstpage}

\maketitle

\begin{abstract}{Photoconduction, recombination}

Experiments often show that the photoconductance $\sigma$ of a semiconductor 
system and the light intensity $I$ are related by $\sigma\sim I^\gamma$.
Conventional theories give a satisfactory explanation for $\gamma=1$ or $\gamma=\frac{1}{2}$,
but anamalous exponents close to $\gamma=\frac{3}{4}$ are often observed.
This paper argues that there is a universal anomalous regime for which
$\gamma=\frac{3}{4}$ (or $\gamma=\frac{2}{3}$ in two-dimensional samples), 
resulting from the kinetics of electron-hole recombination being
controlled by Coulombic attraction. Because the local electric fields are 
extremely high, the theory appeals to the \lq hot-carrier' model 
for transport in semiconductors.
 
\end{abstract}

\maketitle

\section{Introduction}
\label{sec: 1}

Many semiconductor devices rely on excess charge carriers which are
either injected or else created by absorbing photons (Sze, 1981). 
Electron-hole recombination is therefore a process which is central to 
undestanding the dynamics of these systems. The radiative and non-radiative
mechanisms for recombination have been extensively studied, starting 
with seminal work of Shockley and Read (1952). 
In this paper I shall argue that there is a universal regime for the kinetics of 
recombination which has not been described in
earlier work. The evidence comes from considering an anomalous
exponent describing the response of  photoconductors, where the 
photoconductance $\sigma$ and the light intensity $I$ are typically 
related by $\sigma\sim I^\gamma$: values close to $\gamma=\frac{3}{4}$
are frequently observed, whereas conventional theory only gives a satisfactory 
account for $\gamma=1$ or $\gamma=\frac{1}{2}$. I argue that 
accounting for the role of Coulomb interaction in the kinetics of electron-hole 
recombination leads to an explantion for the anomalous exponent, $\gamma$.
A version of this model which uses the \lq hot-carrier' concept (discussed in Ziman, 1972),
as opposed to linear response, provides the most satisfying
description and the best fit to experimental data. This model gives 
$\gamma=\frac{3}{4}$ in three dimesnions, and $\gamma=\frac{2}{3}$
for \lq two-dimesnional' systems, where the photoresponsive layer is very shallow.

The photocurrent excited in a semiconductor might be intuitively 
expected to be proportional to the light intensity, and there are 
reasonable models (discussed below) which do imply a linear dependence. 
A simple and plausible argument however leads to
a different conclusion. Consider a simple photocell 
consisting of a semiconductor (undoped) with two electrodes attached. 
Light falling on the device creates electron-hole pairs (at a rate which 
is proportional to the intensity of radiation, $I$). These mobile carriers 
allow the device to conduct electricity, with a conductivity $\sigma$, which is 
proportional to the density of charge carriers. If the density of 
mobile electrons is $n_{\rm e}$ and the density of holes is $n_{\rm h}$,
the conductivity is
\begin{equation}
\label{eq: 1.1}
\sigma=e(D_{\rm e}n_{\rm e}+D_{\rm h}n_{\rm h})
\end{equation}
where $D_{\rm e}$ and $D_{\rm h}$ are the electron and hole diffusion
coefficients, and $e$ is the electron charge. Because the semiconductor 
is assumed to be intrinsic, the number of electrons equals the number of 
holes, so that $n_{\rm e}=n_{\rm h}\equiv n$. The rate of change of the 
excitation density $n$ contains a source term, proprtional to the intensity, 
and a sink term, which represents the rate for electron-hole
recombination: 
\begin{equation}
\label{eq: 1.2}
\frac{{\rm d}n}{{\rm d}t}=AI - f(n)
\end{equation}
where $A$ is a constant and $f(n)$ is the rate for recombination of 
electron-hole pairs at density $n$. It is natural to invoke the 
law of mass action, and to propose that the rate of recombination is 
proportional to the product of the number density of electrons and of holes:
that is the recombination term is 
${\rm d}n_{\rm e}/{\rm d}t={\rm d}n_{\rm h}/{\rm d}t=-Cn_{\rm e}n_{\rm h}$ 
(for some constant $C$), so that $f(n)=Cn^2$. In this case the equilibrium concentration $n$ and 
consequently the conductance are proportional to $\sqrt{I}$.

Thus there appear to be two plausible models for the recombination rate,
which are distinguished by looking at the exponent $\gamma$ for a power-law 
dependence of conductance $G$ upon light intensity:
\begin{equation}
\label{eq: 1.3}
G  \sim I^\gamma
\ .
\end{equation}
The model described above indicates $\gamma=\frac{1}{2}$. The intuitive
expectation that $\gamma=1$ is consistent with a picture where the 
electrons and holes become immobilised on \lq recombination centres',
which exist independently of the illumination. 

It is not immediately clear which value of $\gamma$ is correct, and it
is necessary to turn to experiment to decide the issue. Both values,
$\gamma=1$ and $\gamma=\frac{1}{2}$ are observed in different experiments,
corresponding respectively to mobile charge carriers being destroyed by 
\lq monomolecular' or \lq bimolecular' processes.
But a surprising aspect of the experimental literature is that intermediate
values of $\gamma$ (hereafter termed {\sl anomalous}) are often observed
over several decades of intensity. 

The existence of anomalous exponents has been discussed by 
Rose (1963), but his model is unsatisfying because it assumes that 
there is a set of impurity states in the band gap with a density of states 
which is an exponential function of energy. Careful studies of Lifshitz tails
pioneered by Halperin and Lax (1966) indicate that the density of states has a 
more complex functional dependence, such as an exponential of a power of energy
(measured from the band edge): see Thirumalai (1986). Because of these results
it seems implausible to use an assumption that the density of states is a simple
exponential in order to explain an anomalous response which has a well-defined exponent
over decades of intensity. The model described by Rose is also 
vulnerable to a more fundamental objection, namely that it relies on an artificial
demarcation between electron states which are traps and those which are recombination
centres. The theory proposed by Rose indicates that 
$\gamma $ should have a monotonic variation with temperature. 
This hypothesis is not compatible with experimental observations by Wronski and Daniel (1981)
where the temperature was varied over a wide range: the exponent 
of the anomalous regime was found to be insensitive to temperature (which is
consistent with the \lq universality' hypothesis developed in this paper).

Reading the literature suggests that where anaomalous exponents are reported,
they are frequently close to $\frac{3}{4}$, particularly in cases where 
a convincing fit to a power-law is demonstrated over two or more 
decades of intensity. The works which I have become aware of are discussed 
in section \ref{sec: 2}. Allowing for errors in fitting the exponent, the experimental results
suggest that $\gamma=\frac{3}{4}$ could be a \lq universal' exponent for 
the anomalous regime.

The samples used in experiments where the anamalous response is 
observed often require specialist equipment for their preparation, and 
the reported values of $\gamma$ are not always consistent. 
In view of the very surprising nature of the anomalous photoconductive response, 
it is desirable to characterise the effect in a readily accessible system. 
As well as reviewing the experimental literature, section \ref{sec: 2} 
will discuss the intensity dependence of the conductance of some commercially 
available photocells. 
An anomalous photoconductive response with $\gamma=\frac{3}{4}$ was observed
in three of the four samples. The remaining one showed an anomalous regime with 
$\gamma=\frac{2}{3}$.

Section \ref{sec: 3} discusses the theory for the rate of recombination.
It is pointed out that the law of \lq mass action' assumes implicitly that
electrons and holes find each other ($\gamma=\frac{1}{2}$ case)
or a recombination trap ($\gamma=1$ case) by diffusion.
Because the electrons and holes are electrically charged, there 
is a regime in which their attraction 
ensures that they collide more rapidly than if their motion was undirected 
diffusion. I show how this leads to a model for the anomalous exponent
$\gamma$. The value of $\gamma$ predicted by this model depends
upon how the carrier drift velocity $v$ varies with the electric field ${\cal E}$.
The natural assumption is to apply linear response theory, and to assume
that $v\propto {\cal E}$. However, the electric field due to the electrostatic 
attraction of the carriers is extremely large. It is known that for very large electric fields the motion
of charge carriers becomes very insensitive to the electric field
(Ziman, 1972). This is accounted for by assuming $v\propto {\cal E}^\beta$,
with $\beta$ very small. In the limit as $\beta\to 0$, the expression for the 
anomalous exponent gives $\gamma=\frac{3}{4}$ in three dimensions and
$\gamma=\frac{2}{3}$ in two dimensions.

\section{Experimental evidence}
\label{sec: 2}

\subsection{Discussion of earlier investigations}
\label{sec: 2.2}

Anomalous photoconductive response has been observed in a wide
variety of semiconductors and semiconductor based devices.  
The results which I have found reported are summarised in table 
\ref{tab: 1}. Some authors (Bakr, 2002, Kaplan and Kaplan, 1998, 2002, 
Spear {\em et al}, 1974) report a broad range of values of $\gamma$,
with the exponent depending upon temperature, or other parameters. 
In the remaining cases where either a single value of $\gamma $ or a narrow range is reported, its value 
appears to lie in the range $0.7-0.8$, apart from three \lq outliers' 
which are discussed below. Small deviations of the exponent fitted in 
experiments can result from various effects, such as the limits of the intensity range extending
into regions where the anomalous regime is breaking down, or the dielectric 
properties of the materials being modified by the concentration of mobile 
carriers. For this reason, it is unlikely that differences in 
these values of $\gamma$ are significant. The experimental literature
is consistent with the view that there could be a robust regime with 
an exponent close to the middle of the range $0.7-0.8$,
while different exponents may be observable in other systems.

\begin{table}
\caption{Summary of previously published anomalous exponents, $\gamma$.}
\begin{tabular}{llcc}\hline
\label{tab: 1}
Sourrce                            &      Material                                               &   $\gamma$      &  Remarks  \\ \hline       
Arene \& Baixeras (1984)   & $a{\rm -Si\,H}$                                         & $0.78$               & 4 decades \\
Arene \& Baixeras (1984)    & $a{\rm -Si\,H}$                                       & $0.55$               &  3.5 decades \\
Bakr (2002)                        &${\rm As}_2{\rm Sb}_3$ film                    &$0.58-0.71$         & 1.5 decades \\
Jie {\em et al} (2006)          & ${\rm Cd\,S}$ nanowires                          & $0.74$-$0.77$   & 1.5 decades \\
Kaplan \& Kaplan (1998)    &${\rm As}_2{\rm Se}_3$                            &$0.44-0.86$                   &2 decades, freq. varied\\
Kaplan \& Kaplan (2002)    &$a{\rm -Se}$                                             &$0.61$-$0.96$     & 2 decades, freq. varied \\ 
Kind {\em et al} (2002)      &  ${\rm Zn}\,{\rm O}$ nanowires                 & $0.8$                & 4 decades \\
quoted in Rose (1963)        & ${\rm Sb}_2{\rm S}_3$ (Vidicon cameras)  & $0.68$              &\lq  several decades' \\
Spear {\em et al}  (1974)  &$a{\rm -Si}$                                               &$0.54$-$0.9$      &method unclear\\
Wronski \& Daniel  (1981)   & $a{\rm -Si\,H}$                                & $0.83$               &  2-3 decades, T varied \\
Wronski \& Daniel (1981)   & $a{\rm -Si\,H}$                                 &$0.7$                  & 3 decades \\
\hline
\end{tabular}
\end{table}

There are three \lq outliers', where a single value of $\gamma $ is quoted
which is not in the range $0.7-0.8$. Wronski and Daniel show results which are 
consistent with $\gamma=0.83$ over a range of temperatures. 
Rose (1963) reports that Vidicon camera tubes 
(which used ${\rm Sb}_2{\rm S}_3$) give $\gamma=0.68$ over \lq several decades'. This value is sufficiently
far from the other exponents that it might have a different origin. Also Arene and Baixeras (1984)  
show a plot with $\gamma=0.55$ over nearly four decades. In this case,
the exponent is so close to the bimolecular case $\gamma=\frac{1}{2}$ that
it is questionable whether this is an anomalous response. 

The theory presented by Rose (1963) predicts that $\gamma$ has the following dependence 
upon the temperature $T$:
\begin{equation}
\label{eq: 2.1}
\gamma=\left\{ 
\begin{array}{cc}
\frac{T_0}{T+T_0} & T\le T_0 \\
\frac{1}{2}            & T>T_0
\end{array}
\right .
\end{equation}
where $T_0$ is an effective temperature characterising the exponential density 
of states in the band gap. A more recent paper developing this approach is 
Schellenberg and Kao (1988).
Wronski and Daniel (1981) examined the intensity dependence 
of the photocurrent in hydrogenated amorphous silicon at a variety off 
temperatures ranging from $147{\rm K}$ to $267{\rm K}$.
Figure 5 of their paper demonstrates that the exponent $\gamma $ is close to
$0.83$ and independent of $T$ at lower intensities, crossing over to $0.5$ at higher intensity.
Their data are, therefore, incompatible with (\ref{eq: 2.1}).

There are some published results which contradict the evidence in
Wronski and Daniel (1981), but these works do not present a 
coherent picture.
Papers by Spear {\em et al} (1974), Kaplan and Kaplan (1998)
and Bakr (2002) report 
that $\gamma $ does depend upon temperature, but their exponents vary erratically and do not show agreement
with (\ref{eq: 2.1}). It is not clear how the intensity dependence of the exponents
obtained by Spear {\em et al} were extracted from the experimental data.

In table \ref{tab: 1} I have not included reports (such as Bube, 1957, Bube and Lind, 1958) 
where anomalous photoconductive response was observed but where the samples displayed non-Ohmic behaviour. 

\subsection{Investigation on commercial photocells}
\label{sec: 2.1}

In discussing the earlier experimental work I have pointed out that much of the the 
published experimental data are consistent with a previously un-remarked universality,
where $\gamma\approx \frac{3}{4}$. Some of the other literature suggests, however, that the 
values of the anamalous exponent are highly variable and may be erratic and hard to reproduce. By definition, universal
behaviour does not shy away from observation. It might be observable in a simple
experiment with a generic semiconductor sample. With this motivation, I measured
the response of four different commercially available light-dependent resistors at room 
temperature ($14-17^\circ {\rm C}$). The photocells, samples 1-4, were obtained from Maplin Electronics.
Their catalogue numbers were: sample 1 - N56AY , sample 2 - N57AY , 
sample 3 - N53AY , sample 4 - N46AY. The specifications do not
include the composition of the sensitive semiconductor layer, but that is not
relevant to the question of whether the anamolous exponent is universal.

\begin{figure}
\centerline{\includegraphics[width=14.0cm]{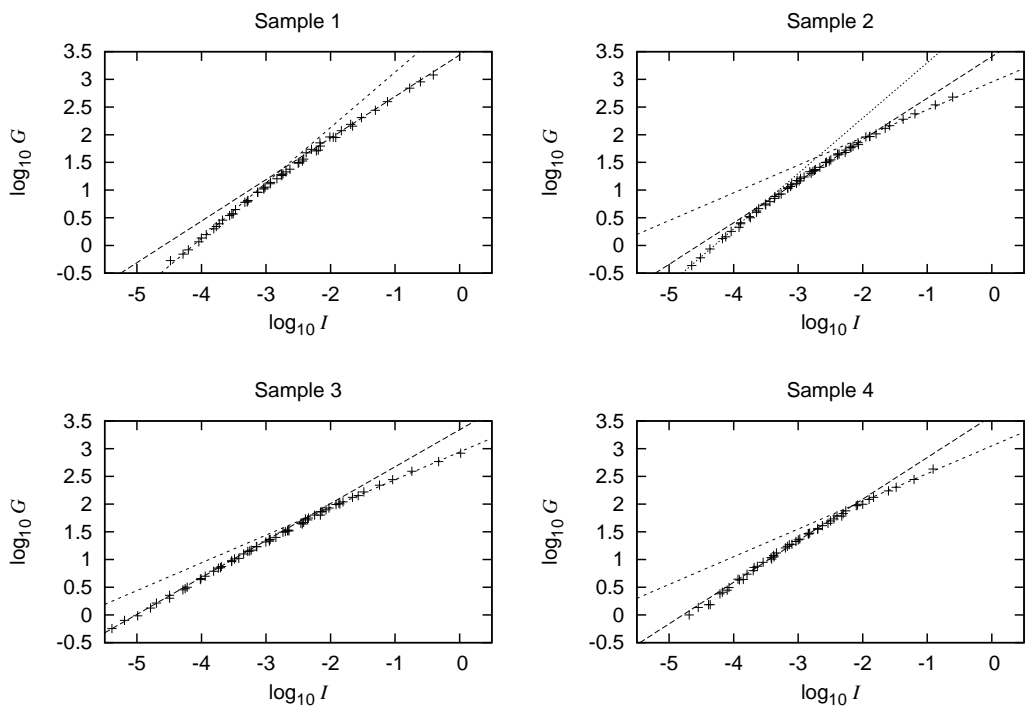}}
\caption{\label{fig: 1} Plot of $\log_{10}G$ (the conductance $G$ measured
in units of $10^{-6}\Omega^{-1}$) versus $\log_{10}I$ (with the intensity $I$ in arbitrary 
units). Each sample showed an anomalous regime. The slopes of the linear fits are as follows. Sample 1: $\gamma=1,\frac{3}{4}$, 
Sample 2: $\gamma=1,\frac{3}{4},\frac{1}{2}$, Sample 3: $\gamma=\frac{2}{3},\frac{1}{2}$, 
Sample 4: $\gamma=\frac{3}{4},\frac{1}{2}$.}
\end{figure}

The intensity of light from a Tungsten bulb was reduced by stopping down 
the aperture of a light box, and by increasing the distance between the aperture and
the photocell. Some of the higher intensity data used an unshielded $15\,{\rm W}$ bulb, varying the distance
from the sample. The response of each photocell was found to be Ohmic 
for both low and high light intensities, and its dark 
current was negligible. All of the data plotted in figure \ref{fig: 1} used a potential
of $3\,{\rm V}$.

The intensity was varied over more than four decades. Figure \ref{fig: 1} shows logarithmic plots of 
conductance $G$ (in units of $10^{-6}\Omega^{-1}$) versus intensity $I$. The light intensity scale is in arbitrary units, chosen so that
the $15\,{\rm W}$ pearlescent bulb at $1\,{\rm m}$ gave an intensity $I=4\times 10^{-3}$. For each 
sample there was clear evidence for an anomalous regime. Sample 1 shows a monomolecular regime
($\gamma=1$) at low intensity, crossing over to an anomalous regime at higher intensity for which
$\gamma=\frac{3}{4}$ gives an excellent fit over two decades. Sample 2 showed an anomalous regime
with $\gamma=\frac{3}{4}$ spanning nearly two decades at intermediate intensities between regimes with $\gamma=1$ and
$\gamma=\frac{1}{2}$. Sample 3 showed an anomalous regime for which $\gamma=\frac{2}{3}$
provides an excellent fit over three decades, crossing over to bimolecular behaviour ($\gamma=\frac{1}{2}$) 
at high intensities. Sample 4 shows an anomalous rigime for which $\gamma=\frac{3}{4}$ provides a good
fit, crossing over to $\gamma=\frac{1}{2}$ at high intensity. 

Three out of four samples showed anomalous behaviour with $\gamma=\frac{3}{4}$, providing strong
support for the hypothesis that this is a \lq universal' phenomenon. It is noteworthy that the other sample 
produced an exponent very close to that reported by Rose (1963) for the Vidicon camera.  All four 
samples showed crossover to other regimes, similar to that seen in figure 5 of Wronski and Daniel (1981).

\section{Theory}
\label{sec: 3}

\subsection{A critique of the law of mass action and an alternative theory} 
\label{sec: 3.1} 

My own experiments and much of the previously published work is consistent the view the there are  
three different regimes which can be observed in different systems, or over 
different ranges of intensity within the same system. In the order where
they might be expected to be observed as the light intensity is increased, 
these are: 

\begin{enumerate}

\item {\sl Monomolecular regime}. Here $\gamma=1$, which is consistent with 
carriers being trapped on a fixed number of recombination centres.
In the low-intensity limit it is expected that a carrier will always
encounter such a trap before encountering a carrier with the opposite sign.

\item {\sl Anomalous regime}. The origin
of this will be considered below. In cases where a crossover has been
observed as intensity is varied, this regime appears to occur in between the 
monomolecular and bimolecular regimes.

\item {\sl Bimolecular regime}. Here $\gamma=\frac{1}{2}$, which is consistent with 
carriers being removed by electron-hole recombination at a rate proprtional
to the product of the electron and hole concentrations.

\end{enumerate}

As the terms {\sl monomolecular} and {\sl bimolecular} suggest, 
the exponents $\gamma=1$ and $\gamma=\frac{1}{2}$ are both explained by 
invoking the law of mass action, as outined in the introduction. The law of mass action is applicable 
in situations where the reactive species move randomly by diffusion, without 
any interaction occurring until they are in contact. This idealisation is 
realistic for chemical reactions in which there is no electrostatic interaction. 
It may also be valid when the species are charged but when the rate 
constant for the chemical reaction is small, so the species reach a local quasi-equilibrium.

In the case of electron-hole recombination, there is Coulombic attraction 
between electrons and holes. This can pull them together, so that the 
time required for a collisions to occur may be less than it would be for 
randomly diffusing particles. 
The effect of recombination events on the density $n$ may be written ${\rm d}n/{\rm d}t=n/\tau$, where 
$n$ is the number of either species, and $\tau$ is the lifetime for an electron
or hole to survive without collisions. The following discussion uses scaling 
arguments rather than detailed calculations of the electron dynamics,
but these arguments are sufficient to make a precise determination
of the anomalous exponent. 

If the collisions occur due to random diffusion of electrons and holes,
then the lifetime $\tau$ will be inversely proportional to the density of the other species,
so that ${\rm d}n/{\rm d}t\propto n^2$. If the Coulombic force is significant, 
the timescale for collision will be the time required for an electron and a hole 
to be pulled together by their mutual electrostatic attraction, and $\tau$ may have a 
different dependence upon the carrier density, $n$.
In this case  $\tau$ is expected to be proportional
to the characteristic distance $L$ that an electron and a hole have to travel in 
order to make a collision. The timescale is also inversely proportional to the 
characteristic velocity $v$ for relative motion of the electrons and the holes:
$\tau \sim L/v$. If the density of carriers is $n$, we clearly have $L \sim n^{-1/d}$,
where $d$ is the effective spatial dimension of the photosensitive region penetrated by the radiation.
(That is, we set $d=3$, unless the depth is small compared to $L$,
when we would use $d=2$). 
The velocity $v$ will depend upon the local electric field ${\cal E}$ driving the drift towards collision.
The charges $\pm e$ of the electrons and holes are independent of the density, but the 
lengthscale $L$ depends upon the carrier density so that ${\cal E}\sim L^{-1}\sim n^{1/d}$.

In order to determine how $\tau$ depends upon $n$, it remains to specify how the typical
drift velocity $v$ depends upon the electric field ${\cal E}$. There are two possibilities:

\begin{enumerate}

\item At first sight it seems natural to invoke linear response theory, and 
to propose that $v\propto {\cal E}$. It should, however, be noted that because the 
lengthscale $L$ is very small, the electric field ${\cal E}$ will be very large. For example, 
the electric field due to an electron at a distance of $10^{-8}\,{\rm m}$ 
is of order $10^7\,{\rm V}\,{\rm m}^{-1}$. These electric fields are so high that 
the applicability of linear response theory is questionable.

\item It is argued that \lq hot-carriers' in semiconductors have a damping rate
due to interaction with phonons which rises very steeply with energy (Ziman 1972). 
As a consequence, the velocity of mobile 
carriers in semiconductors becomes insensitive to the applied electric 
field, and may even appproach a limiting value as the electric field increases (Ferry 1975, Arora 1984).
These hot carrier effects are considered to be significant for electric fields
in excess of $10^5\,{\rm V}\,{\rm m}^{-1}$. 

\end{enumerate}

The physics underlying the hot carrier model will be discussed in greater detail
below, after considering its consequences. In the following it will be assumed that 
the relation between carrier velocity and electric field may be described by a power-law, 
such that 
\begin{equation}
\label{eq: 3.1}
v\sim {\cal E}^\beta\ .
\end{equation}
The linear response model corresponds to setting $\beta=1$. The hot carrier model,
where the velocity is quite insensitive to the microscopic electric field, 
is consistent with either assigning a very small value to $\beta$, or else taking the limit 
as $\beta \to 0$ in the final result.
Before discussing the value of $\beta$, let us consider the formula for the anomalous exponent. 
We have a recombination rate of the form
\begin{equation}
\label{eq: 3.2}
\frac{{\rm d}n}{{\rm d}t}\sim -\frac{nv}{L}\sim n n^{1/d} {\cal E}^\beta \sim n^{\left(1+\frac{1}{d}+\frac{\beta}{d}\right)}\ .  
\end{equation}
Since the rate of recombination is proportional to the flux $I$, we find
$n\sim I^\gamma$ with
\begin{equation}
\label{eq: 3.3}
\gamma=\left(1+\frac{1}{d}+\frac{\beta}{d}\right)^{-1}=\frac{d}{d+1+\beta}
\end{equation}
so that this theory is consistent with the photoconductivity having a power-law dependence
upon $I$ with a non-trivial exponent. Equation (\ref{eq: 3.3}) is the central theoretical result of this
paper. In three dimensions the exponent takes the 
value $\gamma=\frac{3}{5}$ for the first model considered above, where the drift velocity of
the carriers is determined by linear response theory. In the second version of the model, 
where the hot-carrier theory is applied and the limit $\beta\to 0$ is taken, the exponent is $\gamma=\frac{3}{4}$, which is 
consistent with most of the experimental results. For $d=2$, the hot carrier model gives 
$\gamma=\frac{2}{3}$, which is consistent with the results for my own sample 3, with the 
exponent quoted by Rose (1963) for Vidicon cameras, and with the mid-range for the results
of Bakr (2002) on ${\rm As}_2{\rm Sb}_3$ films.

It should be noted that while the non-linear \lq hot-carrier' effect influences the timescale
for electron-hole combination, the response to the small externally applied field is 
determined by equation (\ref{eq: 1.1}) and is Ohmic.

Other authors (see Schlangenotto, Meader and Gerlach, 1974, Hangleiter, 1993) have suggested 
that Coulomb interactions can influence the rate of electron-hole recombination, but these works
consider electron-hole correlations which will modify the constant $C$ in the mass-action formula
for the collision rate, ${\rm d}n_{\rm e}/{\rm d}t=C\,n_{\rm e}n_{\rm h}$. Their calculations
are only relevant to the bi-molecular process.

\subsection{Discussion of the hot carrier theory}
\label{sec: 3.3}

Now consider the response of an electron or hole to the local 
electric field. Note that because we consider the microscopic structure of the 
electric field on a scale of the separation of the carriers, the field 
is not screened by polarisation effects. The samples where the
anomalous response is observed are usually amorphous or highly disordered
microcrystalline systems. The following discussion will assume that the carriers
and the phonons are moving in a highly disordered environment.

The energy available from the electrostatic interaction potential is much higher than the 
thermal energy. The electrons and holes can therefore become \lq hot' carriers', because 
their energies relative to the band edge may be much 
higher than the thermal energy scale.

Studies of spectral hole-burning in disordered semiconductors (Hegarty and Sturge, 1985)
indicate that the motion of electrons and holes is strongly damped
by the effects of phonons. Because of the 
energy loss to phonons, the electrons and holes are expected
to remain close to the bottom of the conduction band or to the top of the valence band,
respectively, despite having energies which are higher than the thermal energy. 

The physics of hot-carrier transport is extremely complex, and in the following
discussion issues such as valley degeneracy will be ignored. The objective
is to give a motivation for assigning a small value to $\beta$ in (\ref{eq: 3.1}), based on a 
reasonable but simplified model for the interaction of charge carriers 
and phonons.

The electric field ${\cal E}$ will cause an electron (or hole) to accelerate 
to a velocity $v$ corresponding to an energy $E=\frac{1}{2}mv^2$, where
$m$ is the effective mass of the carrier (we assume that the electrons or 
holes remain close to bottom or top of their bands, where the particle dispersion 
relations are approximately parabolic and the phonon dispersion can be neglected). 
A particle with velocity $v$ will then gain energy from the electric field at a rate 
$\dot E=e{\cal E}v\sim {\cal E}E^{1/2}$. This acceleration is opposed by the 
tendency of the motion of the charged particle to excite phonons: a particle with 
energy $E$ will dissipate energy to phonons at a rate which will be written 
$\dot E=R(E)E$, where $R(E)$ is a rate coefficient. The energy of a carrier
accelerated by a constant electric field is therefore given by the solution 
of the energy balance equation
\begin{equation}
\label{eq: 3.4}
\frac{{\rm d}E}{{\rm d}t}=e{\cal E}\sqrt{2 E/m}-R(E)E\ .
\end{equation}

The hot carrier picture asserts that rate coefficient $R(E)$ increases very rapidly with 
energy $E$. Various arguments and calculations have been advanced to support
this claim (Ziman 1972, Ferry 1975, Arora 1984). In the following I present a simplified but plausible model, which yields 
$R(E)\sim E^{9/2}$. 

The loss of energy by an electron to the phonons can be seen as analogous to
spontaneous emission from an atomic state, with the phonons playing a role
analogous to the photons in electrodynamics. Recall the expression for the 
rate $R_{nm}$ of spontaneous emission from an atomic level with energy $E_n$ to
a level with lower energy, $E_m$:
\begin{equation}
\label{eq: 3.5}
R_{nm}=\frac{4\alpha}{3c^2}|M_{nm}|^2\omega_{nm}^3
\end{equation}
where $\omega_{nm}=(E_n-E_m)/\hbar $ is the frequency of the photon which is 
emitted, and $M_{nm}$ is the corresponding dipole matrix element ($c$ is the 
speed of light and $\alpha$ is the fine structure constant). This expression
can be adapted to describe the spontaneous decay of electronic states by emission of phonons.
For simplicity it will be assumed that the matrix elements $M_{nm}$ for the electron-phonon
coupling in the highly disordered envirnoment are random numbers, with statistics
which are independent of the energies $E_n$ and $E_m$. The rate of emission into
any given state of lower energy is therefore proportional to $\Delta E^3$, where $\Delta E=E_n-E_m$.
The number of available electronic states with energy below $E$ is $N(E)\sim E^{3/2}$. Upon integrating 
over the lower energy, the rate coefficient scales as $R(E)\sim E^{9/2}$.  
The steady-state solution of equation (\ref{eq: 3.4}) is therefore ${\cal E}\sim E^5$, so that the terminal 
velocity is 
\begin{equation}
\label{eq: 3.6}
v\sim {\cal E}^{1/10}
\end{equation}
that is $\beta=\frac{1}{10}$. This supports the assertion of the hot-carrier
model, that the electron or hole drift velocity becomes highly insensitive to the 
electric field. Some authors (for example, Ferry 1975, Arora 1984) have gone further, and have suggested that the 
drift velocity approaches an asymptote as the electric field increases. This corresponds
to setting $\beta=0$ in (\ref{eq: 3.3}).

\section{Conclusions}
\label{sec: 4}

The theoretical understanding of the anomalous photoconductive response has been 
based upon arguments due to Rose (1963), which make a questionable assumption
about the density of trapping states in the band gap. This theory predicts that the 
anomalous exponent $\gamma$ is temperature dependent, in contradiction
with experiments by Wronski and Daniel (1981).

This paper has proposed an alternative explanation of the anomalous response, based
upon the idea that the kinetics of electron-hole recombination are driven by Coulomb
attraction. The recombination time is then determined by the distance carriers
have to travel, and their velocity $v$, which is assumed to have a power-law dependence 
upon the local electric field ${\cal E}$: $v\sim {\cal E}^\beta$. This leads to a simple
expression for the anomalous exponent, equation (\ref{eq: 3.3}). Because 
the electric fields are extremely high it is not appropriate to 
use linear response theory. In the hot-carrier picture for transport 
at high electric fields, the drift velocity has very weak dependence upon the electric field,
which justifies setting $\beta=0$ in equation (\ref{eq: 3.3}). 

This model leads to the prediction that the anomalous regime has universal
exponents of $\gamma=\frac{3}{4}$ (three dimensions) or $\gamma=\frac{2}{3}$
(for thin samples which are effectively two-dimensional). In support of this 
universality hypothesis, I have pointed out that many of the reported values of 
$\gamma$ for the anomalous regime are close to one or other of these numbers.
Furthermore, experiments on four commercially available photocells
showed anomalous regimes with $\gamma=\frac{3}{4}$ in three cases,
and $\gamma=\frac{2}{3}$ in the remaining case.

{\em Acknowledgments.} I thank Doron Cohen of Ben Gurion University 
for bringing
some references on anomalous photoconductive response to my attention.

\end{document}